\documentclass[a4paper]{jpconf}
\usepackage{lmodern}
\usepackage{lmodern}

\usepackage[LGR,T1]{fontenc}
\usepackage[latin9]{inputenc}
\usepackage{geometry}
\usepackage{amsmath}
\usepackage{amssymb}
\usepackage{setspace}
\usepackage{esint}
\makeatletter

\DeclareRobustCommand{\greektext}{%
  \fontencoding{LGR}\selectfont\def\encodingdefault{LGR}}
\DeclareRobustCommand{\textgreek}[1]{\leavevmode{\greektext #1}}
\DeclareFontEncoding{LGR}{}{}
\DeclareTextSymbol{\~}{LGR}{126}

\usepackage{slashed}

\usepackage{amsthm}\usepackage{amsfonts}\usepackage[british]{babel}
\usepackage{slashed}\usepackage{afterpage}\usepackage{cite}
\usepackage{pdfsync}

\renewcommand{\thefootnote}{\fnsymbol{footnote}}

\def\appendix#1{\addtocounter{section}{1}\setcounter{equation}{0}
\renewcommand{\thesection}{\Alph{section}}
\section*{
\thesection\protect\indent \parbox[t]{11.715cm} {#1}}
\addcontentsline{toc}{section}{Appendix\thesection\ \ \ #1} }

\def\be{\begin{equation}}
\def\ee{\end{equation}}
\def\bea{\begin{eqnarray}}
\def\eea{\end{eqnarray}}

\newcommand{\del}{\partial}

\numberwithin{equation}{section}

\usepackage{babel}

\makeatother

\usepackage{babel}
\begin{document}

\title{Noncommutative geometry, Grand Symmetry and twisted spectral triple}

\author{Agostino Devastato}

\address{Dipartimento
di Fisica, Universit\`{a}  di Napoli Federico II}
\address{INFN, Sezione di Napoli, Monte S.~Angelo, Via Cintia, 80126 Napoli, Italy}

\ead{agostino.devastato@na.infn.it ; astinodevastato@gmail.com }

\begin{abstract}
In the noncommutative geometry approach to the standard model we discuss
the possibility to derive the extra scalar field \textgreek{sv} -
initially suggested by particle physicist to stabilize the electroweak
vacuum - from a  ``grand algebra''  that contains the usual standard
model algebra. We introduce the Connes-Moscovici twisted spectral
triples for the Grand Symmetry model, to cure a technical problem,
that is the appearance, together with the field \textgreek{sv}, of
unbounded vectorial terms. The twist makes these terms bounded, and
also permits to understand the breaking making the computation of
the Higgs mass compatible with the 126 GeV experimental value.
\end{abstract}
\textit{Talk given by A. D. at Roma workshop on Conceptual and Technical
Challenges for Quantum Gravity, September 2014}

\section{Introduction}

Noncommutative geometry~\cite{Connesbook,Landi,Ticos,ConnesMarcolli}
allows to study a large variety of geometrical frameworks from a totally
algebraic approach. Particularly, it is very useful in the derivation
of models in high energy physics, for example the Yang-Mills gauge
theories~\cite{ConnesLott,Schucker,AC2M2,Walterreview,Jureit}. The
mathematical structure of noncommutative geometry is based on three
algebraic objects: a $C^{*}-$algebra $\mathcal{A}$, a Hilbert space
$\mathcal{H}$ and a generalization of the Dirac operator $D$ . These
three elements already naturally appear in some elementary applications
of quantum field theory, for instance the harmonic oscillator in which
the set of physical observables (position, time, energy) is an associative
algebra; spinors, describing the system, belong to an Hilbert space,
and the Dirac operator $\slashed{\partial}=-i\gamma^{\mu}\partial_{\mu}$
which determines dynamic. The set of these three elements $(\mathcal{A},\mathcal{H},D)$
is named \textit{spectral triple}. If the triple is supplemented
with two other ingredients, a \textit{chirality operator} $\Gamma$
and an anti-unitary operator $J$ called \textit{real structure},
then the spectral triple is said a \textit{graded-real spectral triple}
$(\mathcal{A},\mathcal{H},D;J,\Gamma).$

In the current state, gauge theories require a noncommutative geometry
structure that is understood to be an \textit{almost commutative}
geometry, i.e. the product of continuous geometry, representing space-time,
times an internal algebra of finite dimensional matrices. Not many
models can be described in this theory because there are a set
of seven mathematical constraints to be respected. For example, the
\textit{oder-zero condition} 
\begin{equation}
\left[a,Jb^{*}J^{-1}\right]=0\,\,,\forall a,b\in\mathcal{A}\label{Order zero}
\end{equation}
ensuring the same action of the algebra on particles and antiparticles;
or the \textit{first-order condition} 
\begin{equation}
\left[\left[D,a\right],Jb^{*}J^{-1}\right]=0\,\,,\forall a,b\in\mathcal{A}\label{First Order}
\end{equation}
which ensures to obtain a fluctuated Dirac operator of order one in
the derivatives. In this geometric framework the spectral action principle~\cite{spectralaction}
enables the description of the full standard model of high energy
physics, including the Higgs mechanism, putting it on the same footing geometrical footing as general relativity and making it possible the unification
with gravity. Moreover, noncommutative geometry allows to deal with
another classical problem of the standard model: the three gauge coupling
constants run with energy, and at energies comprised between $10^{13}-10^{17}GeV$
they are very close, but, in view of present data, they fail to meet
at a single point; a first possible way to overcome this problem is
to enlarge the Hilbert space, adding new fermions \cite{ULP}. However
it is also possible to deduce, from the spectral action expansion,
higher dimensional terms in the Lagrangian, improving the unification,
\cite{Nomadi} without touching the Hilbert space. 

\section{The Grand Symmetry}

Another useful aspect of noncommutative geometry concerns the prediction
of the Higgs mass. This latter, at unification scale, is a function
of the other parameters of the theory, especially the Yukawa coupling
of fermions $y_{f}$ and the value of the unified gauge couplings
$g=g_{3}=g_{2}=\frac{5}{3}g_{1}$. Assuming there is no new physics
between the electroweak and the unification scales, i.e. the \textit{big
desert hypothesis}, the flow of the Higgs mass under the renormalization
group yields a prediction around $170GeV$, \cite{AC2M2}. It is not possible in the noncommutative geometry
approach to the standard model as well as in the usual Weinberg-Salam
electroweak theory, to predict a Higgs mass near to its experimental
value, $m_{H}\simeq126GeV$ \cite{particledata} without incurring problems of instability. There is, in
fact, an instability in the electroweak vacuum which is meta-stable
rather than stable (see \cite{near-critic} for the most recent
update). This inconsistency strongly suggests that something may be
going on. In particular, particle physicists have shown how it is
possible to cure this instability, assuming the existence of a new
scalar field - usually denoted $\sigma$ - suitably coupled to the
Higgs. In \cite{ccforgotafield} the noncommutative
geometry model has been enlarged obtaining the field $\sigma$ by
turning the Majorana coupling constant $y_{R}$ , in the finite dimensional
part of the Dirac operator, into a field: 
\begin{equation}
y_{R}\to y_{R}\sigma(x)
\end{equation}
However, by definition the bosonic fields in noncommutative geometry
are generated by \textit{inner fluctuations} of the Dirac operator,
defining the 1-form connection $A,$ 
\begin{equation}
A:=\sum_{i}a_{i}\left[D,b_{i}\right]\,\,,\mbox{with}\, a_{i},b_{i}\in\mathcal{A}
\end{equation}
and, unfortunately, this is not the case for the field $\sigma$ because
of the first-order condition \ref{First Order} on the Majorana part
of the Dirac operator, forcing the related one-form to be zero:
\begin{equation}
\left[\left[D_{M},a\right],Jb^{*}J^{-1}\right]=0\,\,\,\Longrightarrow\left[D_{M},a\right]=0\,\,\,\Longrightarrow A_{\sigma}=0\,\,.
\end{equation}

A possible way to generate spontaneously $\sigma$ , is to take advantage
of the fermion doubling in the Hilbert space $\mathcal{H}$ of the
standard model \cite{fermiondoubling}, introducing the Grand
Symmetry \cite{DLM}, a model based on a larger symmetry than the usual
one and mixing gauge and spin degrees of freedom. 
The spectral triple of the Grand Symmetry is the same of the standard model apart from the algebra.  It is the product of the spectral triple $(C^\infty(M), L^2
(M, S), \slashed{\del},\gamma^5,\mathcal{J})$ of a compact
Riemannian manifold M $-$ where $L^2
(M, S)$ is the Hilbert space of square integrable spinors over M,  $\slashed{\del}$ is the usual Dirac operator, $\gamma^5$ the product of the four euclidean gamma matrices and $\mathcal{J}$ the usual charge conjugation operator $-$
by a finite dimensional spectral triple:
\begin{align}
\mathcal{H}_{F} & =\mathcal{H}_{R}\oplus\mathcal{H}_{L}\oplus\mathcal{H}_{R}^{c}\oplus\mathcal{H}_{L}^{c}=\mathbb{C}^{32}\,\,\,\nonumber \\
D_{F} & =D_{0}+D_{M}=\left(\begin{array}{cccc}
0 & \mathcal{M} & \mathcal{M}_{R} & 0\\
\mathcal{M}^{\dagger} & 0 & 0 & 0\\
\mathcal{M}_{R}^{\dagger} & 0 & 0 & \mathcal{M}^{*}\\
0 & 0 & \mathcal{M}^{T} & 0
\end{array}\right)\nonumber \\
\gamma_{F} & =\mbox{diag}(\mathbb{I}_{8},-\mathbb{I}_{8},-\mathbb{I}_{8},\mathbb{I}_{8})\nonumber \\
J_{F} & =\left(\begin{array}{cc}
0 & \mathbb{I}_{16}\\
\mathbb{I}_{16} & 0
\end{array}\right)cc\cdot
\end{align}
the finite Hilbert space $\mathcal{H}_F$ is the usual one of the standard model for one particles family; the matrix $\mathcal{M}$ contains the
quarks, leptons and neutrinos Dirac masses with CKM mixing; while the matrix $\mathcal{M}_R$ contains the Majorana neutrinos mass $y_R$ and forms the Majorana part of the Dirac operator, namely $D_M$. 

The choice of the finite algebra $\mathcal{A}_F$ will be more involved. Under natural assumptions on the representation of the algebra, it
is shown in \cite{DLM} that the algebra in the spectral triple of the
standard model should be a sub-algebra of $C^{\infty}(M)\otimes\mathcal{A}_{F}$
with
\begin{equation}
\mathcal{A}_{F}=\mathbb{M}_{a}(\mathbb{H})\oplus\mathbb{M}_{2a}(\mathbb{C})\,\,,\,\,\mbox{with }a\in\mathbb{N}.\label{General ALGEBRA}
\end{equation}
The algebra of the standard model, 
\[
\mathcal{A}_{sm}:=\mathbb{C}\oplus\mathbb{H}\oplus\mathbb{M}_{3}(\mathbb{C})
\]
is obtained from $\mathcal{A}_{F}$ for $a=2$, by using in addition
to \ref{Order zero} and \ref{First Order}, another axiom of the
theory, namely the grading condition, i.e. every element of the algebra
has to commute with the chirality operator, $\left[a,\Gamma\right]=0.$

Differentely, the Grand Symmetry model starts with the ``grand algebra''
$(a=4\,\,\mbox{in}\,\ref{General ALGEBRA})$, 
\begin{equation}
\mathcal{A}_{G}=\mathbb{M}_{4}(\mathbb{H})\oplus\mathbb{M}_{8}(\mathbb{C})
\end{equation}
and one can generate the field $\sigma$ by inner fuctuation, satisfying
the first-order condition imposed by the majorana part $D_{M}$ of
Dirac operator $D$. In fact, by using again the grading and the first-order
conditions one has the following reductions:
\begin{align}
\mathcal{A}_{G}\,\,\underrightarrow{\mbox{grading }}\mathcal{\, A}'_{G} & =\left[\mathbb{M}_{2}(\mathbb{H})_{L}\oplus\mathbb{M}_{2}(\mathbb{H})_{R}\right]\oplus\left[\mathbb{M}_{4}(\mbox{\ensuremath{\mathbb{C}}})_{r}\oplus\mathbb{M}_{4}(\mbox{\ensuremath{\mathbb{C}}})_{l}\right]\\
\mathcal{A}'_{G\,}\underrightarrow{\mbox{1st order}\,}\mathcal{A}''_{G} & =\left(\mbox{\ensuremath{\mathbb{H}}}_{L}\oplus\mbox{\ensuremath{\mathbb{H}}'}_{L}\oplus\mbox{\ensuremath{\mathbb{C}}}_{R}\oplus\mbox{\ensuremath{\mathbb{C}}'}_{R}\right)\oplus\left(\mathbb{C}_{l}\oplus\mathbb{M}_{3}(\mathbb{C})_{l}\oplus\mathbb{C}_{r}\oplus\mathbb{M}_{3}(\mathbb{C})_{r}\right)\nonumber 
\end{align}
with three of the four complex algebras identified $\mbox{\ensuremath{\mathbb{C}}}_{R}=\mbox{\ensuremath{\mathbb{C}}}_{r}=\mbox{\ensuremath{\mathbb{C}}}_{l}.$
The $\sigma$ field will be given by the difference of two elements
of the remaining complex algebras $\mbox{\ensuremath{\mathbb{C}}}_{r}$
and $\mbox{\ensuremath{\mathbb{C}}'}_{R}$:
\begin{equation}
\sigma\sim y_{R}(c_{R}-c'_{R}),\,\mbox{ with}\, c_{R}\in\mbox{\ensuremath{\mathbb{C}}}_{R}\mbox{ and}\, c'_{R}\in\mbox{\ensuremath{\mathbb{C}}'}_{R}
\end{equation}
The final reduction to the standard model, that is $\mbox{\ensuremath{\mathbb{C}}}_{R}=\mbox{\ensuremath{\mathbb{C}}}'_{R}$,
$\mbox{\ensuremath{\mathbb{H}}}_{L}=\mbox{\ensuremath{\mathbb{H}}'}_{L}$
and $\mathbb{M}_{3}(\mathbb{C})_{l}=\mathbb{M}_{3}(\mathbb{C})_{r}$
is obtained by the first-order condition on the free Dirac operator:
\be
\slashed{D}:=\slashed{\partial}\otimes\mathbb{I}_{F}.
\ee

\section{Twisted Grand Symmetry}

If, on one hand, the Grand Symmetry has the advantage to generate
the field $\sigma$ in the same way than the other gauge fields, on
the other hand it contains two weak points. The first limit is that
the breaking to the standard model algebra $\mathcal{A}_{sm}$ is
obtained by a mathematical condition, i.e. the first-order condition
imposed on the free Dirac operator:
\begin{equation}
\mbox{1-st order condition on }\slashed{D}: \,\,\mathcal{\, A}''_{G}\rightarrow\mathcal{A}_{sm}\,.\label{SM breaking}
\end{equation}
Of course, it would be preferable if the reduction to the standard
model was obtained by a dynamical breaking rather than a mathematical
reduction.

The second and more important question concerns the unboundeness of
the one-forms connections. Unfortunately, before the last breaking
\ref{SM breaking} not only is the first-order condition not satisfied,
but the commutator 
\begin{equation}
\left[\slashed{D},a\right]\,,\,\mbox{with}\, a\in C^{\infty}(M)\otimes\mathcal{A}_{G}
\end{equation}
 is never bounded. This is problematic both for physics, because the
connection 1-form containing the gauge bosons is unbounded, and from
a mathematical point of view, because the construction of a Fredholm
module over $\mathcal{A}$ and Hochschild character cocycle depends
on the boundedness of the commutator \cite{Connes:2000mw}.

In \cite{Devastato:2014bta} both the problems have been solved by using a twisted
spectral triple $(\mathcal{A},\mathcal{H},D;\rho)$, introduced by Connes and Moscovici in \cite{Connes:1938fk} to incorporate type III examples, such as those arising from the transverse
geometry of codimension one foliations. Rather than requiring
the boundedness of the commutator, one asks that there exists a automorphism
$\rho$ of $\mathcal{A}$ such that the twisted commutator 
\begin{equation}
\left[\slashed{D},a\right]_{\rho}:=\slashed{D}a-\rho(a)\slashed{D}
\end{equation}
is bounded for any $a\in\mathcal{A}_{G}$. In fact, one can show that
for a suitable choice of a subalgebra $\mathcal{B}\subset C^{\infty}(M)\otimes\mathcal{A}_{G}$,
a twisted fluctuation of $D=\slashed{D}+D_{M}$, satisfying the first
order condition, generates a field $\sigma$ - very similiar to the
one of \cite{DLM}. Together with this new scalar field
one has also the generation of an additional bounded vector field
$X_{\mu}$, whose potential will lead a breaking mechanism to the standard model.

Explicitly, $\mathcal{B}$ is the sub-algebra $\mathbb{H}^{2}\oplus\mathbb{C}^{2}\oplus\mathbb{M}_{3}(\mathbb{C})$
of $\mathcal{A}_{G}$. Labelling the two copies of the quaternions
and complex algebras by the left/right spinorial indices $l,r$ and
the left/right internal indices $L,\, R$, that is
\begin{equation}
\mathcal{B}=\mathbb{H}_{L}^{l}\oplus\mathbb{H}_{L}^{r}\oplus\mathbb{C}_{R}^{l}\oplus\mathbb{C}_{R}^{r}\oplus\mathbb{M}_{3}(\mathbb{C})
\end{equation}
the automorphism $\rho$ consists in the exchange of the left/right
spinorial indices:
\begin{equation}
\rho:\,(\, q_{L}^{l},\, q_{L}^{r},\, c_{R}^{l},\, c_{R}^{r},\, m)\to(q_{L}^{r},\, q_{L}^{l},\, c_{R}^{r},\, c_{R}^{l},\, m)
\end{equation}
where $m\in M_{3}(\mathbb{C})$ while the $c$'s and $q$'s are complex
numbers and quaternions belonging to their respective copy of $\mathbb{C}$
and $\mathbb{H}$.

Furthermore, the breaking to the standard model is now spontaneous,
as conjectured in \cite{DLM}. Namely the reduction of the grand
algebra $\mathcal{A}_{G}$ to $\mathcal{A}_{sm}$ is obtained dynamically,
as a minimum of the potential coming from the spectral action. The
scalar field $\sigma$ then plays a role similar as the Higgs field
in the electroweak symmetry breaking. Precisely, since the standard
model algebra $\mathcal{A}_{sm}$ is the subalgebra of $\mathcal{B}$
invariant under the twist, one can naturally introduce as physical
degrees of freedom the fields
\begin{equation}
\Delta(X_{\mu}):=X_{\mu}-\rho(X_{\mu})\,\,,\,\,\Delta(\sigma):=\left(\sigma-\rho(\sigma)\right)D_{M}\label{new fields}
\end{equation}
 to measure how far the grand symmetry is from the SM. In (\ref{new fields})
$X_{\mu}$ and $\sigma$ are respectively the twisted fluctuations
of $\slashed{D}$ and $D_{M}$. The new potential, coming from the
spectral action expansion, has the form
\begin{equation}
V=V_{X}+V_{\sigma}+V_{X\sigma}
\end{equation}
where $V_{X}$ depends only by the fields $\Delta(X_{\mu})$, $V_{\sigma}$
depends only by the field $\Delta(\sigma)$ and $V_{X\sigma}$ is
an interaction term containing both the fields. In \cite[sect. 5.6]{Devastato:2014bta}
it is shown how the minimum of this potential is obtained for
\begin{align}
\Delta(X_{\mu}) & =0\\
\Delta(\sigma) & =0\nonumber 
\end{align}
which means the invariance of $X_{\mu}$ and $\sigma$ under the twist
$\rho$. This condition is verified for the biggest invariant sub-algebra
of $\mathcal{B}$ i.e. $\mathcal{A}_{sm}$. 

\section{Conclusions}

Let us get the conclusions: the main idea of the \textit{twisted Grand
Symmetry} model is that the scalar field $\sigma$ is related to the
dynamical breaking of the grand symmetry to the standard model. This
idea was already formulated in \cite{DLM}, but it was not fully implemented
because, without the twist, the fluctuation of the free Dirac operator
by the grand algebra $\mathcal{A}_{G}$ yielded an operator whose
square was a non-minimal Laplacian. Almost simultaneously, a similar
idea was implemented in \cite{CCvS}, where the bigger symmetry did not
come from a bigger algebra, but it followed from relaxing the first
order condition. In the twisted case, here presented, we have shown
a reduction mechanism to the standard model very similiar to that of 
\cite{CCvS}. Therefore, it would be interesting to understand the relationship
between the twisted fluctuations and the inner fluctuation without
the first oder condition.
Moreover, the twist $\rho$ is remarkably
  simple, and its mathematical structure should be studied more in
  details, in particular how it should be incorporated in the other axioms
  of noncommutative geometry, for example the orientability condition where the commutator with the Dirac operator has a very important role. Also, the physical meaning of the twist is intriguing:
  the un-twisted version of $\cal B$ forces the action of the algebra to be equal on the right and left components of
  spinors. In this sense the breaking of the grand symmetry to the
  standard model could be interpreted as a sort of "primordial'' chiral
  symmetry breaking.

\ack{The author thanks W. van Suijlekom and F. Lizzi for discussions.}


\begin{thebibliography}{10}

\bibitem{Connesbook} A. Connes, \textit{Noncommutative Geometry},
Academic Press, 1984.

\bibitem{Landi} G. Landi, \textit{An Introduction to Noncommutative
Spaces and their Geometries}, \textsl{Springer Lecture Notes in Physics
51}, Springer Verlag (Berlin Heidelberg) 1997. arXiv:hep-th/9701078.


\bibitem{Ticos} J.M.~Gracia-Bondia, J.C.~Varilly, H.~Figueroa,
\textit{Elements of Noncommutative Geometry}, Birkhauser, 2000.

\bibitem{ConnesMarcolli} A. Connes, M. Marcolli, ``Noncommutative
Geometry, Quantum Fields and Motives'', AMS 2007;

\bibitem{ConnesLott} A.~Connes and J.~Lott, ``Particle Models
And Noncommutative Geometry (expanded Version)'' Nucl.\ Phys.\ Proc.\ Suppl.\ \textbf{18B}
(1991) 29. 


\bibitem{Schucker} T.~Schucker, ``Forces from Connes' geometry,''
Lect.\ Notes Phys.\ \textbf{659} (2005) 285 {[}hep-th/0111236{]}.


\bibitem{AC2M2} A.~H.~Chamseddine, A.~Connes and M.~Marcolli,
``Gravity and the standard model with neutrino mixing'' Adv.\ Theor.\ Math.\ Phys.\ \textbf{11}
(2007) 991 {[}arXiv:hep-th/0610241{]}. 


\bibitem{Walterreview} K.~van den Dungen and W.~D.~van Suijlekom,
``Particle Physics from Almost Commutative Spacetimes'' arXiv:1204.0328
{[}hep-th{]}. 


\bibitem{Jureit} J.~H.~Jureit, T.~Krajewski, T.~Schucker and
C.~A.~Stephan, ``Seesaw and noncommutative geometry'' Phys.\ Lett.\ B
\textbf{654} (2007) 127 {[}arXiv:0801.3731 {[}hep-th{]}{]}. 


\bibitem{spectralaction} A.~H.~Chamseddine and A.~Connes, ``The
spectral action principle,'' Commun.\ Math.\ Phys.\ \textbf{186},
731 (1997) {[}arXiv:hep-th/9606001{]}. 

 \bibitem{ULP} 
 A.~A.~Andrianov, D.~Espriu, M.~A.~Kurkov and F.~Lizzi,
  ``Universal Landau Pole,''
  Phys.\ Rev.\ Lett.\  {\bf 111} (2013) 011601
  [arXiv:1302.4321 [hep-th]].

\bibitem{Nomadi}
 A.~Devastato, F.~Lizzi, C.~V.~Flores, and D.~Vassilevich 
 ``Unification of coupling constants, dimension 6 operators and the spectral action," 
 International Journal of Modern Physics  A, 2015, 30, 1550033
[arXiv:1410.6624  [hep-ph]]

\bibitem{particledata} J. Beringer et al. (Particle Data Group),
Phys. Rev. D86, 010001 (2012).

\bibitem{near-critic}
D.~Buttazzo, G.~Degrassi, P.~P. Giardino, G.~F. Giudice, F.~Sala, and
  A.~Salvio, \emph{Investigating the near-criticality of the {H}iggs boson},
  arXiv:1307.3536 [hep-ph].

\bibitem{ccforgotafield} A.~H.~Chamseddine and A.~Connes, ``Resilience
of the Spectral Standard Model'' arXiv:1208.1030 {[}hep-ph{]}. 



\bibitem{fermiondoubling} F.~Lizzi, G.~Mangano, G.~Miele and G.~Sparano,
  ``Fermion Hilbert space and fermion doubling in the noncommutative geometry approach to gauge theories,''
  Phys.\ Rev.\ D {\bf 55} (1997) 6357
  [hep-th/9610035].

\bibitem{DLM} A.~Devastato, F.~Lizzi and P.~Martinetti, ``Grand
Symmetry, Spectral Action, and the Higgs mass,'' arXiv:1304.0415
{[}hep-th{]}. 

\bibitem{Connes:2000mw}
  A.~Connes and H.~Moscovici,
  ``Cyclic cohomology and Hopf symmetry,''
  math/0002125 [math-oa].

\bibitem{Devastato:2014bta}
  A.~Devastato and P.~Martinetti,
  ``Twisted spectral triple for the Standard Model and spontaneous breaking of the Grand Symmetry,''
  arXiv:1411.1320 [hep-th].

\bibitem{Connes:1938fk}
A.~Connes and H.~Moscovici, \emph{Type {III} and spectral triples}, Traces in
  number theory, geoemtry and quantum fields, Aspects Math. \textbf{E38}
  (2008), no.~Friedt. Vieweg, Wiesbaden, 57--71.


\bibitem{CCvS}
  A.~H.~Chamseddine, A.~Connes and W.~D.~van Suijlekom,
  ``Inner Fluctuations in Noncommutative Geometry without the first order condition,''
  arXiv:1304.7583 [math-ph].




\end{thebibliography}
\end{document}